\documentclass[%
 reprint,
superscriptaddress,
 amsmath,amssymb,
 aps, pra,
]{revtex4-2}

\usepackage{graphicx}
\usepackage[caption=false]{subfig} 
\usepackage{soul}
\usepackage{dcolumn}
\usepackage{comment}
\usepackage{bm}
\usepackage{color}

\usepackage{ulem}

\newcommand{\dd}{\mathrm{d}}

\usepackage[hidelinks]{hyperref}

\begin{document}

\title{Non-equilibrium coexistence between a fluid and a hotter or colder crystal of granular hard disks}
\author{R. Maire}
\affiliation{Universit\'e Paris-Saclay, CNRS, Laboratoire de Physique des Solides, 91405 Orsay, France}
\author{A. Plati}
\affiliation{Universit\'e Paris-Saclay, CNRS, Laboratoire de Physique des Solides, 91405 Orsay, France}
\author{ F. Smallenburg}
\affiliation{Universit\'e Paris-Saclay, CNRS, Laboratoire de Physique des Solides, 91405 Orsay, France}
\author{G. Foffi}\email{giuseppe.foffi@universite-paris-saclay.fr}
\affiliation{Universit\'e Paris-Saclay, CNRS, Laboratoire de Physique des Solides, 91405 Orsay, France}

\date{\today}

\begin{abstract}
   Non-equilibrium phase coexistence is commonly observed in both biological and artificial systems, yet understanding it remains a significant challenge. Unlike equilibrium systems, where free energy provides a unifying framework, the absence of such a quantity in non-equilibrium settings complicates their theoretical understanding. Granular materials, driven out of equilibrium by energy dissipation during collisions, serve as an ideal platform to investigate these systems, offering insights into the parallels and distinctions between equilibrium and non-equilibrium phase behavior. For example, the coexisting dense phase is typically colder than the dilute phase, a result usually attributed to greater dissipation in denser regions. In this article, we demonstrate that this is not always the case. Using a simple numerical granular model, we show that a hot solid and a cold liquid can coexist in granular systems. This counterintuitive phenomenon arises because the collision frequency can be lower in the solid phase than in the liquid phase, consistent with equilibrium results for hard-disk systems. We further demonstrate that kinetic theory can be extended to accurately predict phase temperatures even at very high packing fractions, including within the solid phase. 
   Our results highlight the importance of collisional dynamics and energy exchange in determining phase behavior in granular materials, offering new insights into non-equilibrium phase coexistence and the complex physics underlying granular systems.
\end{abstract}
\maketitle

\section{\label{sec:Intro} Introduction}

The theory of equilibrium phase coexistence as formalized by Gibbs establishes that mechanical, thermal, and chemical equilibrium are necessary conditions for the stability of a heterogeneous substance at equilibrium \cite{gibbs1878equilibrium}. With the recent advancements in non-equilibrium statistical physics, it has become evident that phase coexistence phenomena are equally ubiquitous in out-of-equilibrium systems \cite{cates2015motility, cates2024active, schlogl1972chemical} and that Gibbs' equilibrium conditions must be relaxed for example in active matter systems or in driven chemical phase transitions~\cite{schlogl1972chemical, mandal2019motility}. This realization has sparked significant theoretical and experimental efforts to explore the differences and similarities between these non-equilibrium cases and their well-known equilibrium counterparts \cite{van2019interrupted, cates2010arrested, mani2015effect, prymidis2015self, saha2014clusters, liebchen2015clustering, matas2014hydrodynamic, thutupalli2018flow}. Several peculiarities characterize phase coexistence in non-equilibrium systems. For instance, in active systems, the common tangent construction fails and must be replaced by alternative constructions based on an effective free energy \cite{wittkowski2014scalar}. Likewise, the bulk density of the coexisting phases may depend on the effective surface tension \cite{solon2018generalized, omar2023mechanical}, highlighting the system's deviation from a conventional underlying free energy.  Out of equilibrium, the coarsening dynamics are also unusual, with phenomena such as non-standard roughness of interfaces at coexistence \cite{caballero2024interface,besse2023interface}, peculiar growth of length scales during coarsening \cite{shi2020self, mandal2019motility, pattanayak2021ordering} and reversed Ostwald ripening leading to bubbly phases \cite{fausti2024statistical, tjhung2018cluster} or microphase separation \cite{caporusso2020motility}. While surface tension between phases can still be defined following different equilibrium definitions, these approaches can yield different results in non-equilibrium systems \cite{omar2020microscopic, langford2024mechanics, patch2018curvature, zakine2020surface}. Similarly, macroscopic heat flux between phases and peculiar interfacial properties can be observed in boundary-driven systems \cite{sasa2024non, nakagawa2017liquid,yoshida2024heat, kobayashi2023control, sasa2021stochastic}. Dynamic phenomena such as traveling \cite{ouazan2023self,  fruchart2021non}, pattern-forming \cite{butzhammer2015pattern, ansari2018phase}, and even chasing coexisting phases \cite{chiu2023phase} have been observed as well.

Recently, inspired by biological active systems,  a growing interest in underdamped self-propelled particles surfaced \cite{lowen2020inertial, caprini2020spontaneous, caprini2021inertial,caprini2022role}. Among other things, it has been observed that these particles can undergo a motility-induced phase separation \cite{kuroda2023anomalous,suma2014motility, hecht2022active}, with a  dense phase colder \cite{hecht2024define} than the  dilute one due to reduced effective self-propulsion in the dense phase \cite{mandal2019motility}. This highly non-equilibrium effect was recently observed in an experimental system \cite{caprini2024dynamical}. In contrast, Ref.~\onlinecite{hecht2024motility} reported motility-induced phase separation resulting in a hotter solid. Similar observations, in stark contrast to Gibbs' requirement of thermal equilibrium, were previously made in the study of vibrated granular media, where gas-liquid \cite{risso2018effective, clewett2016minimization, cartes2004van, khain2011hydrodynamics, argentina2002van, herminghaus2017phase, khain2004phase, khain2004oscillatory, brey2016stability, roeller2011liquid, noirhomme2021particle, noirhomme2018threshold} and liquid-solid phase coexistence \cite{komatsu2015roles, clerc2008liquid, luu2013capillarylike, prevost2004nonequilibrium, rivas2012characterization, gotzendorfer2005sublimation, melby2005dynamics,PhysRevLett.81.4369,zuniga2022geometry, reyes2008effect, lobkovsky2009effects} consistently revealed a colder dense phase. These results have traditionally been explained by the assumption that the denser phase must dissipate more energy, and therefore be colder, due to a higher collision frequency compared to the coexisting dilute phase. 

Nonetheless, a hotter solid was recently observed in a driven 2D granular system \cite{gao2024temperature}. This phenomenon was however attributed to boundary effects rather than induced by the inter-particle interaction. In a recent realistic numerical study of a vibrated quasi-2D granular system, we also surprisingly observed that a crystal composed of bidisperse beads can exhibit a hotter solid than the coexisting liquid \cite{plati2024self}. This temperature difference partially arose from geometric effects that cannot occur in a monodisperse system. In this paper, we simplify the previously used model by employing only monodisperse beads and demonstrate that even for such simple systems, it is possible to find a solid phase hotter than the coexisting liquid. Surprisingly, we find that dissipation during collisions can be the key mechanism for a hotter solid. By developing a kinetic theory to explain these results, we show that, although the solid phase is always denser, its particle collision frequency is typically lower than in the coexisting liquid, reducing dissipation and enabling the solid to be hotter.

The paper is organized as follows: in Sec.~\ref{sec: numerical investigations}, we present and simulate a simple 2D granular model in the liquid-solid coexistence region, showing how a hotter solid can emerge. In Sec.~\ref{sec: theory}, we explain this behavior using an equilibrium-like argument based on the collision frequency of hard-disk coexisting phases, and then extend the discussion with a fully non-equilibrium theory. Finally, in Sec.~\ref{sec: Discussion}, we discuss the relevance and broader implications of our findings.

\section{\label{sec: numerical investigations} Numerical investigations}

To explore the temperature differences between the coexisting liquid and solid phases in a system of dissipative hard disks, we consider a simple model system in two distinct limits. In this section, we describe both the model and the simulation methods we use to study them.

\subsection{\label{sec: models} The model}

A quasi-2D vibrated box serves as a paradigmatic system for investigating driven granular systems \cite{maire2024interplay, plati2024quasi,castillo2012fluctuations,castillo2015universality,guzman2018critical,PhysRevLett.81.4369, PhysRevE.59.5855, PhysRevE.70.050301, PhysRevLett.89.044301,castillo2019hyperuniform, reis2006crystallization}. In this setup, particles are confined within a quasi-2D horizontal plane, bounded by parallel top and bottom plates. The system is driven by vertical vibrations of the container, which inject energy into the vertical degrees of freedom of the particles. This energy is subsequently dissipated and redistributed across the horizontal plane through inter-particle collisions. 

To capture the essential physics of the quasi-2D system, we propose the following 2D coarse-grained model. The system is made of 2D dissipative hard disks of mass $m$ and diameter $\sigma$. When two particles collide, they can both dissipate energy and gain momentum, leading to the following collision rule  \cite{brito2013hydrodynamic, maire2024interplay}:
\begin{equation}
    \begin{split}
        \bm v_i'&= \bm v_i - \left[\dfrac{1+\alpha}{2}(\bm v_{ij}\cdot \hat{\bm\sigma}_{ij}) + \Delta \right] \hat{\bm\sigma}_{ij} \\
        \bm v_j'&= \bm v_j + \left[\dfrac{1+\alpha}{2}(\bm v_{ij}\cdot \hat{\bm\sigma}_{ij}) + \Delta \right] \hat{\bm\sigma}_{ij},
    \end{split}
    \label{eq: collRule}
\end{equation}
where $\bm v_i'$ is the post-collisional velocity of particle $i$, while $\boldsymbol{\hat{\sigma}}_{ij}$ and $\bm v_{ij}$ are respectively the unit vector joining particles $i$ and $j$ and the relative velocity between them. The coefficient of restitution $0<\alpha<1$ governs the dissipation during collisions, and $\Delta>0$ is responsible for velocity injection. This velocity injection mimics the energy redistribution during grain-grain collisions in the quasi-2D system described above. The energy change during a collision can be either positive or negative, depending on the relative velocity of the particles. For large relative velocities,  $|\bm v_{ij}|\gg \Delta$ and sufficiently small $\alpha$, the term proportional to $\alpha$ dominates, leading to dissipative collisions. Conversely, at low relative velocities, energy dissipation through $\alpha$ is minimal, and instead, particles gain an additional velocity $\Delta$, effectively injecting energy into the system.

Between collisions, the hard disks are also coupled to an external bath at granular temperature $T_b$ with drag $\gamma$ via a Langevin equation:
\begin{equation}
    \dfrac{\dd \bm v}{\dd t} = -\gamma \bm v + \sqrt{2\gamma T_b/m}\bm{\eta},
    \label{eq: langevin}
\end{equation}
where $\bm \eta$ is a random vectorial white Gaussian noise with unit variance and zero mean. Comparing again to a vibrated quasi-2D granular system, this effective noise would represent the roughness of the top and bottom plates confining the particles in the realistic quasi-2D model. Collisions with these plates cause the hard disks to behave as Brownian particles on the horizontal plane for time scales larger than the one set by the frequency of the shaking \cite{sarracino2010granular, puglisi2014transport}.

We focus on two limiting cases of interest:
\begin{itemize}
    \item The $\mathit{\Delta + \gamma}$ \textit{model}, where $T_b=0$ \cite{maire2024interplay, maire2024Enhancing}. Here, particles lose energy during their free flight according to $v(t)=v_0e^{-\gamma t}$ while collisions, on average, increase the system's energy—though in some cases, they may also lead to energy loss.
    This 2D model is representative of a quasi-2D system in a dynamical regime where particles are shaken with sufficient energy to undergo strong collisions with both the lower and the upper plate. $\gamma$ effectively accounts for the horizontal energy loss due to tangential friction with the smooth plates, while $\Delta$ reproduces the energy transfer between the vertical and horizontal degrees of freedom at collision. The equilibrium limit corresponds to $\Delta\to0$, $\alpha\to1$ and $\gamma\to0$. 
    \item The \textit{granular Langevin model} (GLM), where $\Delta = 0$ \cite{sarracino2010granular, plati2024self}. In this case, all the energy is supplied by the bath, and collisions only dissipate energy. In the quasi-2D system, this corresponds to grains confined between rough plate that are not strongly shaken, which limits their vertical motion and interaction with the upper plate. In this regime, surface asperities transfers energy primarily into the horizontal plane during grain-plate collisions, while grain-grain collisions contribute minimally to the energy change in the horizontal plane. In the limit $\alpha \rightarrow 1$ we recover an equilibrium system.
\end{itemize}  
\begin{figure*}[!t]
\includegraphics[width=0.98\textwidth,clip=true]{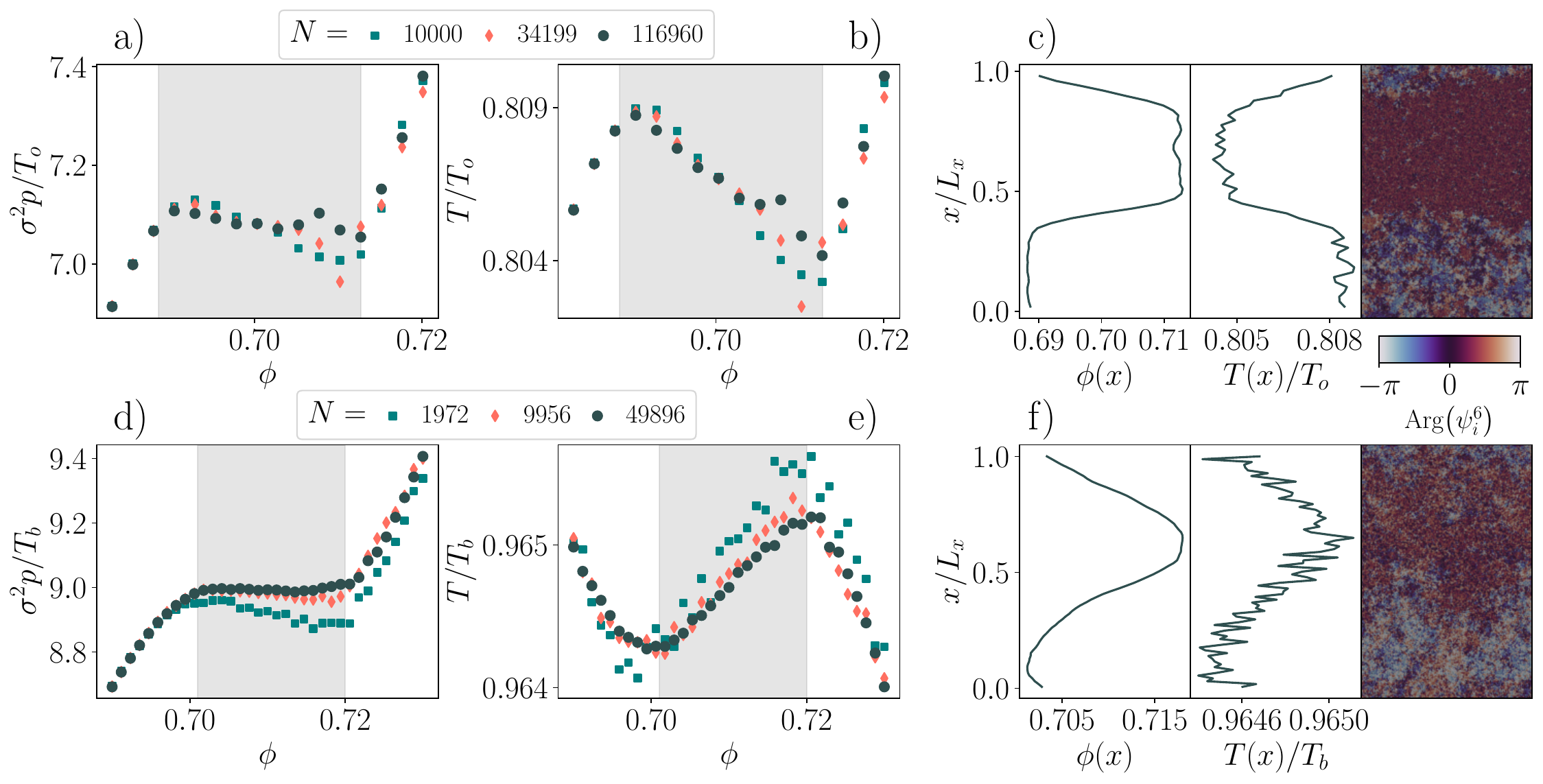} 
\centering
\caption{Top: $\Delta+\gamma$ model ($\Delta/\sigma\gamma=1.1$ and $\alpha = 0.95$), bottom: GLM ($T_b/m(\sigma \gamma)^2=0.125$ and $\alpha = 0.99$). a) and d): Pressure as a function of the density. b) and e):  Energy as a function of the density. First two panels of c) and f): Granular temperature and density averaged over a  $y$ slab in $[x, x+10\sigma]$. Last panel of c) and f): Snapshot of a liquid-solid coexistence, particles are colored by the angle of hexatic local order parameter $\text{Arg}(\psi_i^6)$ with $\psi_i^6 = \sum_{j=1}^{N_i} e^{i6\theta_{ij}}/N_i$, with $N_i$ the number of neighbors to particle $i$ and $\theta_{ij}$ the angle in real space between $i$ and $j$. The density and temperature profiles in c) represent time-averaged values over $2\times 10^4$ measurements of the properties obtained from the time evolution of the displayed snapshot. The measurement window is sufficiently small to ensure that large variations in the density field remain negligible. The coexistence behavior of the GLM is more challenging to visualize compared to that of the $\Delta + \gamma$ model. This difficulty arises because its behavior closely mirrors the equilibrium liquid-hexatic phase coexistence, where the transition between the two phases is extremely weak, resulting in strongly fluctuating interfaces \cite{bernard2011two}. For this system, the profiles are obtained by averaging $10^5$ snapshots over 60 independent realizations. Before performing the averaging, all particles are shifted such that the center of mass \cite{bai2008calculating} is positioned at $x/L_x=0.75$ ensuring that the solid phase consistently resides in the region $0.5<x/L_x<1$ \cite{maggi2021universality, siebert2018critical}. Therefore, unlike in the $\Delta+\gamma$ model, the snapshot of the GLM in f) does not correspond to the ones used to extract the profiles. Note finally that for the $\Delta+\gamma$ model, the lack of a thermal bath means there is no obvious external reference temperature to compare the measured $T$ to. Hence, for this system, we plot the temperature in terms of $T_o$, which we define as the theoretical temperature of the Delta model assuming velocity distributed as a Gaussian and in the absence of drag ($\gamma= 0$). It is given by \cite{brito2013hydrodynamic} $T_o=\frac{1}{2}m\Delta^2\left(\alpha\sqrt{\pi}/2 + \sqrt{\alpha^2 \pi/4+ (1-\alpha)^2}\right)^2/(1-\alpha^2)^2$ and is independent on the density.}  \label{fig: big recap}
\end{figure*}
\subsection{Numerical simulations \label{sec: numerical simulations}}

Using event-driven molecular dynamics \cite{smallenburg2022efficient} we simulate both models described above using up to $N\simeq10^5$ particles in boxes of size $L_x\times L_y$ with periodic boundary conditions. We define $\phi=N\pi (\sigma/2)^2/L_xL_y$ as the packing fraction of the system and focus on the region near the equilibrium liquid-hexatic transition. For simplicity, we will refer to the crystalline or hexatic phase in coexistence with the liquid as the ``solid phase" of the system without making a distinction between the two.

The granular temperature $T$ of the system is defined as \cite{baldassarri2005temperature, puglisi2017temperature}:
\begin{equation}
    T = \frac{1}{2}m\sum_{i=1}^N\langle\bm{v}_i^2\rangle,
    \label{eq: definition temperature}
\end{equation}
where $\langle\dots\rangle$ represents a time average. When explicitly stated, it also includes an average over different initial conditions. Additionally, we define the pressure $p$ as the virial pressure (see Appendix.~\ref{sec: appendix pressure}):

\begin{equation}
    p = \dfrac{NT}{L_xL_y}+\dfrac{m\sigma}{2L_xL_yt}\sum_{\text{coll-ij}}\left(\dfrac{1+\alpha}{2}\left(\bm{v}_{ij}\cdot\hat{\bm{\sigma}}_{ij}\right)+\Delta\right).
\end{equation}

We present the results of our simulations in the coexistence region of the liquid-solid phase transition in Fig.~\ref{fig: big recap}. The top panels show the results for the $\Delta+\gamma$ model, while the bottom panels display the results for the GLM. For the $\Delta+\gamma$ model, we observe a Mayer-Wood loop \cite{bernard2011two, mayer1965interfacial, engel2013hard} in the pressure (Panel a), indicating a first-order phase transition. More intriguingly, in the coexistence region, the granular temperature also exhibits a non-monotonic trend (Panel b). Unlike the loop in pressure, this phenomenon is not related to interfacial effects, as it does not disappear with increasing $N$. Instead, it is most likely associated with the coexistence of two phases at different temperatures. Indeed, if the existence of a lever-rule for the density field is assumed \cite{plati2024self}, the temperature field which is slaved to the density and crystallization field should also follow a lever-rule. As the density increases, a larger fraction of the system transitions to the solid phase, following the conventional lever-rule for the density field. Consequently, a greater portion of the system attains the temperature of the solid. Since in Panel b the global temperature decreases, we hypothesize that the coexisting solid phase is colder than the coexisting liquid phase. This is confirmed by direct computations of the temperature $T(x)$ and density $\phi(x)$ profiles, corresponding to a given phase separation, averaged over multiple realizations and snapshots (Panels c). For the GLM, the pressure loop is still evident (Panel d), indicating a first-order phase transition. The temperature also presents the expected non-monotonic trend (Panel e), but in this case, the solid is hotter than the liquid, as the global temperature increases. This observation is corroborated by direct computations of the local temperature and densities (Panels f). 

Therefore, we find a solid that is hotter than the liquid in the GLM, even though the collisions are purely dissipative. Surprisingly, in the $\Delta+\gamma$ model, where energy is injected during collisions, the solid is colder.

Before delving into the theory explaining these intriguing results, we note that, both for the GLM and the $\mathit{\Delta+\gamma}$ model, the energy difference between the two phases is rather low (i.e. between 0.1 and 0.3 $\%$). This rather small temperature difference  ultimately stems from the small density difference between the liquid and the solid, as well as our choice of parameters, which keeps the system relatively close to equilibrium. Maintaining this proximity to equilibrium is essential for the validity of the theoretical framework we will develop. Additionally, we point out that in the GLM, further increasing dissipation readily leads to a continuous phase transition and the disappearance of the phase coexistence. This phenomenon is discussed in greater detail in Appendix~\ref{sec: appendix continuous}.

\section{Theory \label{sec: theory}}
\subsection{Equilibrium argument for the temperature difference}

In order to gain intuition about the surprising results concerning the temperature of both phases, we provide an intuitive argument grounded on an equilibrium description. Assuming the system remains close to equilibrium, we will for now utilize equilibrium results for the pressure to infer the possible temperature difference between the two phases that are slightly out of equilibrium.

Consider a coexistence between a liquid and a solid of hard disks \textit{at equilibrium}. For hard disks, the pressure can be directly related to the contact value $g^+$ of the pair correlation function \cite{li2022hard}:
\begin{equation}
    p^{eq}(\phi, T, \phi g^+)=\frac{4\phi}{\sigma^2\pi} T (1 + 2\phi g^+).
\end{equation}
Note that even in equilibrium, we still define $T$ as the average kinetic energy of a particle (see Eq.~\eqref{eq: definition temperature}).
The constraint of mechanical equilibrium between the solid and the liquid phase can therefore be written as:
\begin{equation}
    \phi_s T (1 + 2\phi_s g^+_s) = \phi_l T(1 + 2\phi_l g^+_l),
    \label{eq: equilibrium pressure}
\end{equation}
where $g^+_s$ and $g^+_l$ are the equilibrium pair correlation function at contact in the solid and in the liquid respectively. Note that the temperature $T$ is the same for both phases, as we are considering an equilibrium system. Since $\phi_s > \phi_l$, it follows from Eqs.~\eqref{eq: equilibrium pressure} that:
\begin{equation}
    \phi_s g^+_s < \phi_l g^+_l.
    \label{eq: frank sugestion}
\end{equation}
Moreover, the pressure of a hard-disk system is directly connected to the rate of collisions \cite{li2022hard}, leading to the Enskog expression for the collision frequency $\omega$ \cite{pagonabarraga2001randomly}:
\begin{equation}
    \omega(T, \phi g^+)=8\phi g^+\sqrt{T/\sigma^2\pi m},
    \label{eq: equilibrium freq}
\end{equation}
which holds exactly in equilibrium.
We will later check the validity of this assumption, especially in the solid. 
This implies from Eq.~\eqref{eq: frank sugestion} and \eqref{eq: equilibrium freq} that the collision frequency in the solid is lower than in the liquid at coexistence: \begin{equation}
    \omega(T, \phi_s g_s^+) < \omega(T, \phi_l g_l^+).
    \label{eq: equilibrium collision frequency inequality} 
\end{equation}
In other words, the increase in density in the solid is compensated by a decrease in the collision frequency, such that the momentum exchange rate—or equivalently, the pressure—remains the same in both the liquid and the solid phases.
We now assume that, close to equilibrium, the collision frequency in the coexisting liquid is still higher than in the solid. Under this approximation, in systems where energy is dissipated during collisions, the liquid phase is expected to be colder than in the coexisting solid. The same argument suggests a higher temperature in the coexisting liquid than in the solid when energy is injected during collisions. However, outside the coexistence region, the collision rate increases monotonically with density, causing temperature to decrease with density for dissipative collisions and increase when energy is injected at collision. This contrast between inside and outside the coexistence region explains the non-monotonic trend of the energy in both models shown in Fig.~\ref{fig: big recap}.
\\
In this argument, we assumed that the non-equilibrium systems we consider here can be regarded as small perturbations to an equilibrium system, such that $g^+$ and the temperature of the system are only weakly affected by the introduction of non-equilibrium effects. Although this equilibrium picture provides an intuitive argument for the expected behavior of the temperature difference, it leads to an internal inconsistency, where we use the assumption of thermal equilibrium to show the emergence of a temperature difference.
In the next section, we explicitly take the temperature difference between the two phases into account in our analysis to derive it in a consistent way.

\subsection{Non-equilibrium derivation of the temperature difference}

We now take into account the non-equilibrium nature of our system. Assuming that velocities follow a Gaussian distribution and are uncorrelated before collisions, we find that the pressure $p$ in either phase can be written as (see Appendix \ref{sec: appendix pressure}):
\begin{equation}
    p(\phi, T, \phi g^+)= \frac{4\phi}{\sigma^2\pi} \left[T + \phi g^+\left((1+\alpha)T+ 2\Delta\sqrt{\pi m T}\right)\right],
    \label{eq: pressure averaged}
\end{equation}
where we still find the ideal contribution and the virial contribution proportional to $(1 + \alpha)T$ which represents momentum redistribution due to particle interactions. A new term proportional to $\Delta$ emerges from non-equilibrium velocity injection during collisions \cite{garzo2018enskog}.

The steady state temperature arises from a balance between dissipation and energy injection \cite{brito2013hydrodynamic}. Under the same assumptions as for the pressure, it can be shown that (see Appendix \ref{sec: appendix temperature}):
\begin{equation}
    \begin{split}
        \dfrac{\omega (T, \phi g^+)}{2}&\left(m\Delta^ 2 + \alpha\Delta \sqrt{\pi m T}-(1-\alpha^2) T\right)\\
        &  - 2\gamma (T - T_b)=0.
    \end{split}
    \label{eq: temperature averaged}
\end{equation}
The first term of Eq.~\eqref{eq: temperature averaged} represents the rate of energy change due to collisions, with the expression in parentheses describing the average energy change per collision. The second term accounts for energy exchange with the thermal bath. A crucial consequence of Eq.~\eqref{eq: temperature averaged} is that the temperature depends on the density only through the frequency of collision. Therefore (far from the solid-liquid interface and when $\gamma \neq 0$) the temperatures of the coexisting phases can be assumed to depend solely on their respective values of $\phi g^+$. 

\begin{figure}[t]
\includegraphics[width=0.98\columnwidth,clip=true]{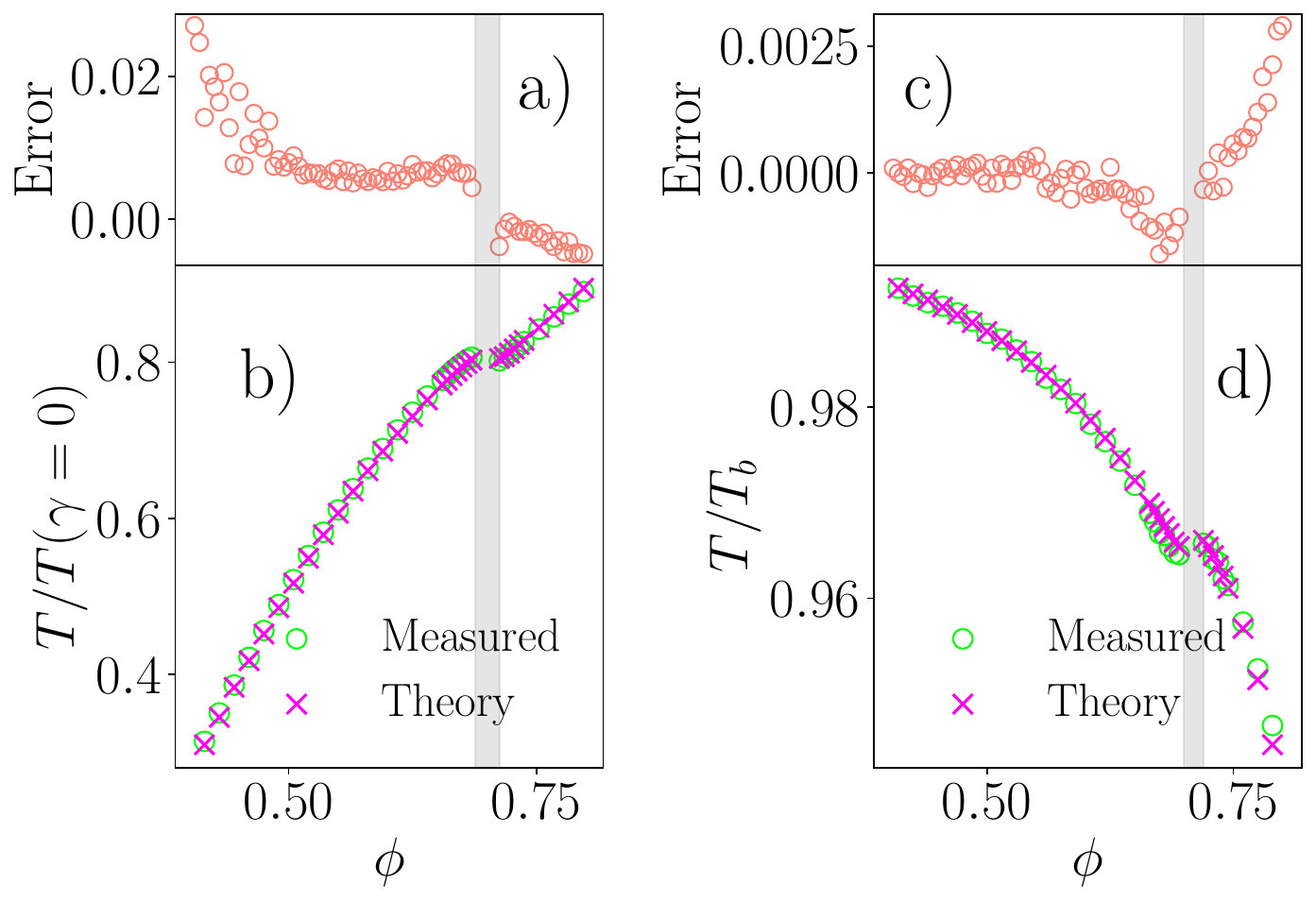}
\centering
\caption{Comparison between kinetic theory predictions and measured values of temperature. Panels a) and b) show the temperature predicted by kinetic theory (using measured values of $g^+$) compared to the directly measured temperature for the $\Delta + \gamma$ model. The Error is calculated as $(T_\text{measured}-T_\text{theory})/T_\text{theory}$. Panels c) and d) present the same comparison for the GLM. Both models are simulated utilizing the same parameters as in Fig.\ref{fig: big recap}.}  \label{fig: check kinetic}
\end{figure}

This approach assumes that the velocity distribution remains approximately Gaussian and that velocities are uncorrelated during collisions. Together, these assumptions lead directly to Enskog’s formula for the collision frequency (Eq.~\eqref{eq: equilibrium freq}), which we apply here. However, since we are interested in dense fluids and solids that, in principle, lie beyond the domain where Enskog theory applies, it is essential to test these assumptions. To this end, in Fig.~\ref{fig: check kinetic}a) and c), we compare measured temperatures to theoretical predictions for both the $\Delta+\gamma$ model and the GLM, respectively. These predictions were obtained using direct measurements of $g^+$ and the numerically calculated root of Eq.~\eqref{eq: temperature averaged}. Overall, the theory agrees well with numerical data, as demonstrated by the errors, see Fig.~\ref{fig: check kinetic}b) and d). Notably, for the $\Delta+\gamma$ model, the theory underperforms in the dilute limit due to an absorbing phase transition near $\phi \simeq 0.25$, where molecular chaos significantly breaks down \cite{maire2024interplay, neel2014dynamics}. In contrast, the GLM behaves as expected: the theory is accurate at low densities but becomes less reliable as the packing fraction is increased. It may seem surprising that Enskog’s theory, often considered valid only in the dilute limit, performs well here, even under non-equilibrium steady states. For instance, the theory is known to inaccurately predict transport coefficients, such as viscosity, in dense fluids \cite{hansen2013theory} as these predictions are highly sensitive to collective effects and long-time tails \cite{hansen2013theory} that Enskog’s framework cannot capture. However, quantities such as the collision frequency that depend solely on the local environment are less affected by these non-local features. Based on these results, we conclude that the kinetic theory can be trusted to accurately predict temperatures within the parameter region studied ($\alpha>0.95$). 
\\
The results presented in Fig.~\ref{fig: check kinetic}, show that the theory predicts well the non-monotonicity of the temperature. However, it requires measuring the values of $g^+$. While equilibrium predictions \cite{hansen2013theory} proved to be good approximations for driven dissipative systems \cite{maire2024interplay,gradenigo2011fluctuating, barrat2002molecular, van1999randomly}, we will demonstrate that we can make useful predictions about the temperatures of both phases in the coexistence region of non-equilibrium hard disk systems without resorting to this approximation or relying on any equation of state.
\\
Having tested the validity of our approximations as a first step, we note that mechanical stability still requires the equality of pressure of both phases:
\begin{equation}
    p(\phi_s, T_s, \phi_s g^+_s) = p(\phi_l, T_l, \phi_l g^+_l).
    \label{eq: equality of pressure}
\end{equation}
For simplicity, we assumed flat interfaces between the phases, avoiding the need to account for Laplace pressure and other interfacial stresses \cite{de2024statistical}, which could influence the temperatures of each phase.
The key insight is that energy change at collision implies that the density dependence of energy arises solely from the collision frequency, which in the Enskog approximation depends on density only through $\phi g^+$ (Eq.~\eqref{eq: equilibrium freq}). This means that $\phi g^+$ can be found if the temperature is known. Indeed, we can isolate $\phi g^+$ in the equation for the temperature Eq.~\eqref{eq: temperature averaged} by inserting $\omega$ defined in Eq.~\eqref{eq: equilibrium freq} into it:
\begin{equation}
    \phi g^+\equiv\mathcal{G}(T)= \dfrac{\sigma \gamma \sqrt{\pi m} (T - T_b)}{2\sqrt{T}(m\Delta^ 2 + \alpha\Delta \sqrt{\pi m T}-(1-\alpha^2) T)}.
    \label{eq: def of G}
\end{equation}
This allows us to eliminate the dependence on $\phi g^+$ in the pressure Eq.~\eqref{eq: equality of pressure}:
\begin{equation}
    p(\phi_s, T_s, \mathcal G(T_s)) = p(\phi_l, T_l, \mathcal G(T_l)).
    \label{eq: equality of pressure reduced}
\end{equation}
By eliminating $\phi g^+$, the pressure now explicitly depends only linearly on the density:
\begin{equation}
    p(\phi, T, \mathcal{G}(T))=\phi \tilde p(T, \mathcal G(T))
\end{equation}
where $\tilde{p}$ is defined as:
\begin{equation}
    \tilde p(T, \mathcal{G}(T)) = \dfrac{\sigma^2 \pi}{4}\frac{p}{\phi},
    \label{eq: definition reduced p}
\end{equation}
which does not depend on $\phi$. Finally, using mechanical equilibrium (see Eq.~\eqref{eq: equality of pressure reduced}) between phases and $\phi_s > \phi_l$, we find a non-equilibrium criterion:
\begin{equation}
    \tilde p(T_s, \mathcal G(T_s)) < \tilde p(T_l, \mathcal G(T_l)).
    \label{eq: criterion pressure}
\end{equation}
Eq.~\eqref{eq: criterion pressure} allows us to determine which phase is hotter based on $\mathcal{G}$, without needing to know an equation of state or $g^+$. For instance, if $\tilde{p}(T)$ is a continuous and increasing function of $T$, then Eq.~\eqref{eq: criterion pressure} implies $T_s < T_l$. However, if $\tilde{p}$ is non-monotonic, the temperature comparison becomes more complex.

\subsection{Hotter or Colder solid?}

We first focus on the $\Delta+\gamma$ model ($T_b = 0$). In this model, $\tilde{p}$ is a continuous and monotonically increasing function of the physically accessible granular temperatures (see Appendix~\ref{sec: appendix details delta}). Consequently, from Eq.~\eqref{eq: criterion pressure}, we find that:
\begin{equation}
    T_s^{\Delta+\gamma} < T_l^{\Delta+\gamma}.
    \label{eq: prediction temperature delta model}
\end{equation}
As confirmed by our simulations, the temperature of the solid is predicted to be lower than that of the liquid, regardless of the parameters $\Delta > 0$ and $\gamma > 0$. Note that this theory still assumes molecular chaos and a Gaussian velocity distribution, therefore it should only be trusted near equilibrium conditions.

The theory for the GLM ($\Delta = 0$) is more intricate than that of the $\Delta+\gamma$ model. For this model, we obtain (see Appendix \ref{sec: appendix details langevin}):
\begin{equation}
    \tilde p^\text{GLM}(\tilde T) = T_b \tilde T \left(1 + \Lambda (1 - \tilde T)\tilde T^{-3/2}  \right),
    \label{eq: p tilde langevin}
\end{equation}
with:
\begin{equation}
    \tilde T = T/T_b \text{ ~~ and ~~  } \Lambda = \frac{\gamma\sigma\sqrt{\pi m/T_b}}{2(1-\alpha)}>0.
\end{equation}
$\Lambda$ is a dimensionless parameter such that, at the Gaussian level, $\Lambda \to \infty$ corresponds to the equilibrium limit. Since the thermal bath is the only source of energy injection, the physical temperatures are constrained by $\tilde{T} < 1$.

For $\Lambda > 1$, $\tilde p$ is continuous and decreasing for $\tilde T < 1$, leading to the result:
\begin{equation}
    T_l^{\text{GLM}} < T_s^{\text{GLM}} \text{ when } \Lambda>1.
    \label{eq: prediction temperature langevin model}
\end{equation}
This explains the outcome observed in Fig.~\ref{fig: big recap}, where the solid was found to be hotter than the liquid in the GLM (with $\Lambda \simeq 10^2$).

However, in the strongly non-equilibrium case where $\Lambda < 1$, unlike the $\Delta+\gamma$ model, $\tilde{p}$ is not monotonically increasing, complicating the analysis. Nevertheless, we saw in Appendix~\ref{sec: appendix continuous} that for these small $\Lambda$ the liquid-solid transition was most likely always continuous (i.e. without phase coexistence).

\section{Discussion \label{sec: Discussion}}

Our investigations have shown that, in a granular system, the solid phase in a solid-liquid coexistence can indeed be hotter than the liquid. A heuristic, equilibrium-based argument for this counterintuitive result is that, at coexistence, the collision frequency in the solid phase is lower than in the liquid phase. Therefore, in systems where energy is dissipated through collisions, it is natural to expect the solid to be hotter than the liquid, even if the former is denser than the latter. 

Although these heuristic arguments provide an intuitive explanation for our findings, they start from an equilibrium assumption, with constant temperature throughout the system, rendering them inconsistent when used to predict temperature differences. The more rigorous analysis in Sec.~\ref{sec: theory} overcomes this limitation by explicitly incorporating the temperature difference between the two phases.

In developing the theory, we made certain assumptions that may not strictly hold out of equilibrium. Specifically, we assumed molecular chaos and a Gaussian velocity distribution, which can easily break down \cite{brilliantov2004kinetic, van1998velocity, esipov1997granular, garzo2018enskog, baxter2007experimental}, particularly in the solid phase. However, we show that close to equilibrium, the predictions from kinetic theory works exceptionally well even in the solid. We also showed in Ref.~\onlinecite{plati2024self} that the theory still works even in a binary mixture where energy equipartition between small and large disks does not hold.  Extensions beyond the Gaussian assumption are possible \cite{gomez2024diffusion, garzo2018enskog, brilliantov2004kinetic, brilliantov2006breakdown} but fall outside the scope of this article.
\\
The models we introduced are simplified compared to realistic granular experiments, such as quasi-2D vibrated granular matter. As a first step, the analysis performed on the two limiting models can be extended to the full system with $\Delta \neq 0$ and $T_b \neq 0$. However, nothing too surprising emerges from this extension: either the bath or $\Delta$ dominates the energy injection, and the temperature difference follows.
In more realistic systems of quasi-2D vibrated monodisperse granular beads, the dense phase was always found to be colder than the dilute one due to effects not accounted for in our simplified 2D model: bistability of the
grain-plate dynamics \cite{PhysRevLett.81.4369,Geminard2003}, strong vertical confinement 
\cite{prevost2004nonequilibrium}, dynamical instabilities \cite{cartes2004van,risso2018effective}, non-homogeneous energy injection \cite{noirhomme2021particle} among others. While our study has the merit of showing that the temperature difference commonly observed in granular system cannot be solely explained by dissipative collision, a crucial next step would be to incorporate these factors into our model to gain a better understanding of what governs the temperature difference between the two phases. Conversely, from an experimental standpoint, it would be interesting to optimize the realistic parameters of the quasi-2D system to get as close as possible to the effective GLM, and to investigate whether it is feasible to obtain a monodisperse granular solid that is hotter than its coexisting liquid. We also note that in multi-component hard disk mixtures, where a variety of crystal and quasi-crystal structures can form \cite{plati2024quasi,fayen2020infinite, fayen2023self, fayen2024quasicrystal}, a granular solid was observed to be hotter than the liquid in a realistic quasi-2D vibrated system. This is because collisions between some species are geometrically impossible in a perfect crystal but not in the liquid, reducing a source of dissipation only in the dense phase \cite{plati2024self}. 
\\
In active matter, motility-induced phase separation can be realized by self-propelled macroscopic agents undergoing dissipative collisions \cite{caprini2024dynamical, van2019interrupted}. Since the phase separation in these systems arises from a dynamical instability associated with a reduction in effective self-propulsion—and consequently in effective temperature—as the system's density increases, it is unlikely that the effect we have identified plays a significant role in these cases.
\\
Additionally, our theory assumes the solid-liquid transition is discontinuous. However, for the GLM, we found that this is not always the case (see Appendix~\ref{sec: appendix continuous}). This could be attributed to the softening of the effective interaction due to dissipative collisions \cite{rodriguez2019granular, bordallo2009effective, velazquez2016effective}. Indeed, particles with potentials softer than hard-core repulsion are known to undergo the standard two steps Kosterlitz-Thouless-Halperin-Nelson-Young melting transition \cite{kosterlitz2018ordering, halperin1978theory, nelson1979dislocation, young1979melting, berezinskii1971destruction} from a solid to a hexatic and then from a hexatic to a liquid phase, which both occur continuously \cite{toledano2021melting, kapfer2015two}. Lastly, the question of hexatic order in these systems is of significant interest and was not addressed in this study. Recent work has shown that crystals formed in the $\Delta+\gamma$ model exhibit translational long-range order due to hyperuniformity \cite{maire2024Enhancing,galliano2023two, kuroda2024long, keta2024longrangeordertwodimensionalsystems}, in striking violation of the Mermin-Wagner theorem. It would be interesting to explore the impact of temperature difference and translational long-range order on the nature of the melting of granular crystals. 

\begin{acknowledgments}

AP acknowledges the Agence Nationale de la Recherche (ANR), Grant ANR-21-CE06-0039, which provided funding for this research. The data that support the findings of this study are available from the corresponding author upon reasonable request.

\end{acknowledgments}

\appendix

\section{Determination of the pressure\label{sec: appendix pressure}}

The typical method for obtaining the pressure at equilibrium is through the virial pressure formula. Out of equilibrium, pressure can either be derived from the Boltzmann equation or through a method analogous to that used in equilibrium. In this work, we follow the latter approach. The pressure $p$ in 2D systems interacting through 2-body forces can be written as \cite{soto2001statistical}:

\begin{equation}
    p = \frac{4 \phi}{ \sigma^2\pi}\left(T - \frac{1}{2N}\sum_{i< j}^N \left\langle \bm{r}_{ij}\cdot \bm {F}_{ij}\right\rangle\right),
    \label{eq: pressure virial}
\end{equation}
where $\bm F_{ij}$ is the force between particles $i$ and $j$.  Similar to equilibrium systems, the Langevin thermostat only contributes to the ideal part of the pressure and does not affect the virial term, which arises from momentum exchanges between particles. Notably, Eq.~\eqref{eq: pressure virial} can also be derived from a direct evaluation of the momentum exchange at a boundary.

The force between particles $i$ and $j$ can be derived from the collision rule Eq.~\eqref{eq: collRule} using $m\dd \bm v_i/\dd t = \bm F_{ij}$:

\begin{equation}
    \bm F_{ij} = -m\left(\dfrac{1 + \alpha}{2}|\bm v_{ij} \cdot \hat{\bm  \sigma}_{ij} | + \Delta\right)\hat{\bm \sigma}_{ij}\delta(t - t^{\text{coll}}_{ij}),
    \label{eq: force}
\end{equation}
with $t^{\text{coll}}_{ij}$ the time of collision between particle $i$ and $j$. Eq.~\eqref{eq: pressure virial} can be simplified using Enskog's collision frequency \eqref{eq: freq}, the expression of the instantaneous force \eqref{eq: force} and the collisional average defined in Eq.~\eqref{eq: MeanValues}:
\begin{equation}
    \begin{split}
    p &= \frac{4 \phi}{ \sigma^2\pi}\left[T + \frac{m\sigma \omega(T, \phi g^+)}{4}\left\langle \dfrac{1 + \alpha}{2}|\bm v_{ij} \cdot \hat{\bm  \sigma}_{ij} | + \Delta\right\rangle_\text{coll}\right]\\
    &=\frac{4\phi}{\sigma^2\pi} \left[T + \phi g^+\sqrt{T}\left((1+\alpha)\sqrt{T}+ 2\sqrt{\pi m}\Delta\right)\right],
    \end{split}
    \label{eq: pressure virial coll average}
\end{equation}
which coincides with Eq.~\eqref{eq: pressure averaged} in the main text.

Note that from the average we can also obtain an expression for the microscopic running pressure used in the Event-Driven simulations to measure the pressure:

\begin{equation}
    p_\text{micro} = \dfrac{NT}{L_xL_y}+\dfrac{m\sigma}{2L_xL_yt}\sum_{\text{coll-ij}}\left(\dfrac{1+\alpha}{2}\left(\bm{v}_{ij}\cdot\hat{\bm{\sigma}}_{ij}\right)+\Delta\right),
\end{equation}
where $t$ is the time simulation window and the sum is over all collisions.

\section{From discontinuous to continuous liquid-solid phase transition\label{sec: appendix continuous}}

\begin{figure*}[t]
  \centering
  \subfloat[Pressure]{ \includegraphics[width=0.48\linewidth]{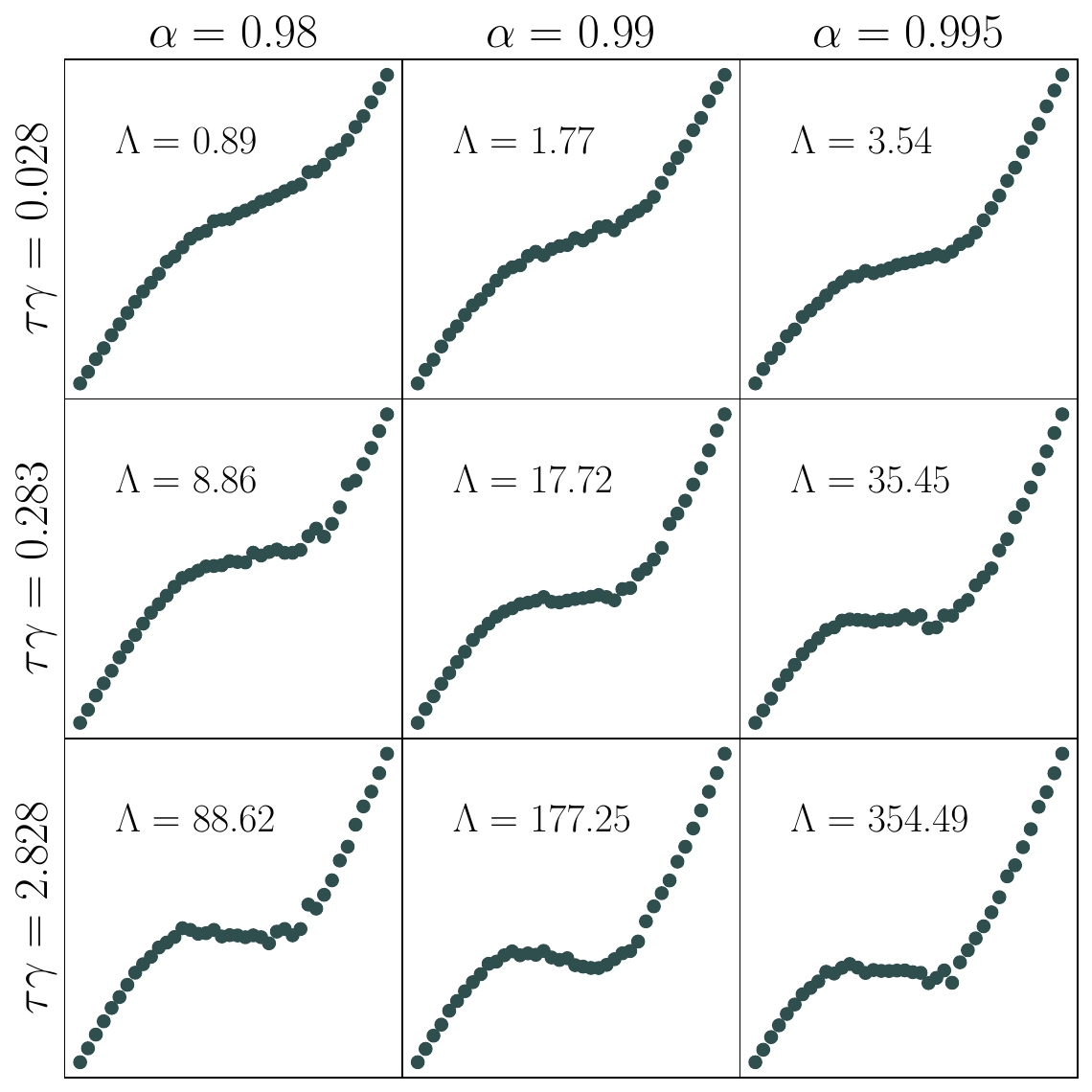}}
  \hfill
  \subfloat[Energy]{ \includegraphics[width=0.48\linewidth]{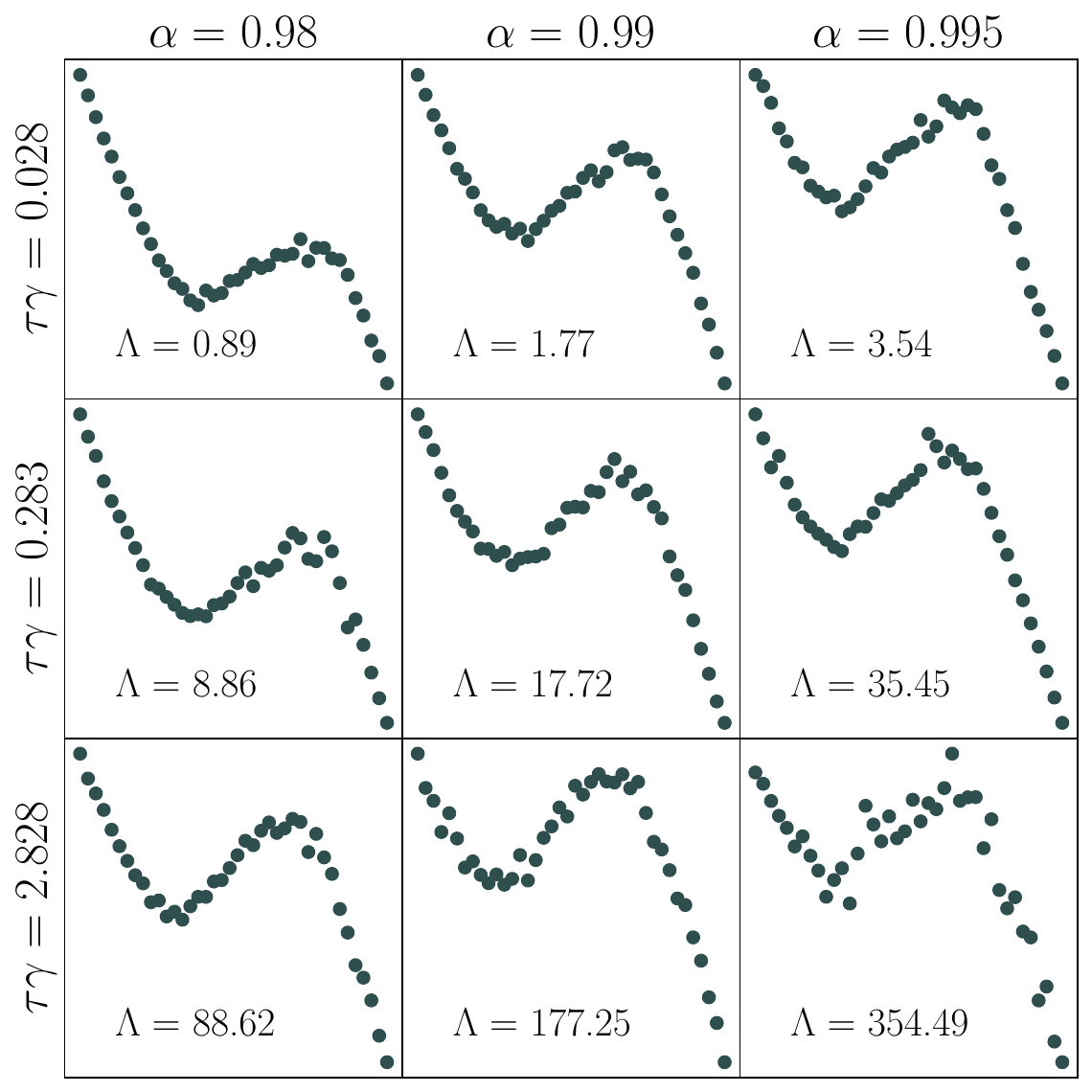}}
  \caption{a) Evolution of pressure (in arbitrary units) as a function of density for the GLM, for various $\gamma$ and $\alpha$ ($\tau = \sqrt{m\sigma^2/T_b}$).
    b) Same as in a), but for the energy instead of the pressure (in arbitrary units). $N = 10^4$.} 
  \label{fig: pressure and energy}
\end{figure*}

The liquid-solid phase transition in the GLM can become continuous under certain conditions. In the equilibrium limit $\Lambda \to \infty$, the model appears to transition from the liquid to the solid phase (most likely an hexatic phase) in a discontinuous manner, as shown in Fig.~\ref{fig: pressure and energy}a). This is consistent with the known behavior of equilibrium hard-disks \cite{bernard2011two}. However, as $\Lambda$ decreases and dissipation becomes more significant, the transition appears to become continuous, with no Mayer-Wood loop \cite{mayer1965interfacial}—indicative of strong, finite-sized interfacial effects—observed in the pressure. This behavior is reminiscent of the classical two-step Kosterlitz-Thouless-Halperin-Nelson-Young (KTHNY) transition, where the liquid-to-hexatic transition is continuous. It has been shown that systems interacting with soft potentials —for example, $V(r) \propto r^{n}$ with $n<6$ — can follow the KTHNY scenario in transitioning from the solid to the liquid phase \cite{toledano2021melting, kapfer2015two}.

In our system, the observed change in behavior as dissipation increases may be explained by an effective softening of the hard-disk potential \cite{rodriguez2019granular, bordallo2009effective, velazquez2016effective}. However, further studies are necessary to fully understand the mechanisms underlying this change in the nature of the transition.

Interestingly, a loop in the energy is consistently observed in Fig.~\ref{fig: pressure and energy}b), regardless of whether the transition is continuous or discontinuous. This suggests that even in the absence of a first-order phase transition, the structural changes associated with the transition to the solid phase lead to an increase in temperature around the transition point.

We finally note that a change from continuous to discontinuous in the liquid-solid transition was also observed in an experimental vibrated granular gas in Ref.~\cite{castillo2012fluctuations, castillo2015universality} upon changing driving parameters and in Ref.~\cite{komatsu2015roles} while changing the properties of the grains.

\section{Determination of the temperature\label{sec: appendix temperature}}

We define the instantaneous energy of the system as $E=m/2\sum_{i=1}^N\bm{v}_i^2$. From the collision rule, the energy change during a collision is given by:
\begin{equation}
    \Delta E = m\Delta^2 + \alpha m(\bm v_{ij}\cdot \hat{\bm\sigma}_{ij})\Delta- m\dfrac{1-\alpha^2}{4}(\bm v_{ij}\cdot \hat{\bm\sigma}_{ij})^2.
    \label{eq: energy change microscopic}
\end{equation}
The evolution of the temperature in the system is then given by:
\begin{equation}
    \dfrac{\dd T}{\dd t}=\frac{\omega}{2}\langle \Delta E\rangle_{\text{coll}} - 2\gamma (T - T_b),
\end{equation}
where $\omega$ is the collision frequency and $\langle\dots\rangle_{\text{coll}}$ is an average over collisions defined from the Boltzmann equation as:
\begin{widetext}
    \begin{equation}
    \langle A(\bm v_{ij}, \hat{\bm\sigma}_{ij})\rangle_{\text{coll}} = \dfrac{\displaystyle{\int d \bm v_i\int d \bm v_j \int d  \hat{\bm\sigma}_{ij} \Theta(-\bm v_{ij}\cdot \hat{\bm\sigma}_{ij})|\bm v_{ij}\cdot  \hat{\bm\sigma}_{ij}| A(\bm v_{ij}, \hat{\bm\sigma}_{ij}) f^{(2)}(\bm v_i, \bm v_j, \hat{\bm\sigma}_{ij})}}{\displaystyle{\int d \bm v_i\int d \bm v_j \int d \hat{\bm\sigma}_{ij} \Theta(-\bm v_{ij}\cdot \hat{\bm\sigma}_{ij})|\bm v_{ij}\cdot \hat{\bm\sigma}_{ij}| f^{(2)}(\bm v_i, \bm v_j, \hat{\bm\sigma}_{ij})}}.
    \label{eq: monstruosity}
    \end{equation}
\end{widetext}
$\Theta$  is the Heaviside function that ensures only particles approaching each other are considered,  $|\bm v_{ij}\cdot  \hat{\bm\sigma}_{ij}|$ can be interpreted as a flux times a cross-section for the hard-disks interaction, and $f^{(2)}(\bm v_1, \bm v_2, \bm\sigma_{ij})$ is the pair distribution function of the velocities. Assuming molecular chaos with Enskog's extension, we simplify the two points velocity distribution:
\begin{equation}
    f^{(2)}(\bm v_i, \bm v_j, \hat{\bm\sigma}_{ij}) \simeq g^+ f(\bm v_i)f(\bm v_j),
\end{equation}
where $g^+$ is the pair correlation function at contact and $f(\bm v)$ the distribution of velocity assumed to be a gaussian. These assumptions of molecular chaos and gaussianity of the velocity distribution let us approximate Eq. \ref{eq: monstruosity} to the following form in 2D:
\begin{equation}
    \langle A\rangle_{\text{coll}} = \dfrac{\displaystyle{\int_{0}^{\infty}\mathrm{d} v\int_{-\pi/2}^{\pi/2}\mathrm{d}\theta\cos(\theta) A(v, \theta) \dfrac{m v^2}{2T}e^{-\dfrac{1}{2}\dfrac{m v^2}{2T}}}}{2\sqrt{\dfrac{\pi T}{m}}}.
    \label{eq: MeanValues}
\end{equation}
These assumptions also lead to the Enskog expression for the collision frequency that we used in the main text:
\begin{equation}
    \omega(T, \phi g^+) = \langle |\bm v |\rangle/l(\phi)= 8 \phi g^+\sqrt{T/(\pi m)}/\sigma.
    \label{eq: freq}
\end{equation}
Eq.~\eqref{eq: freq} can be equivalently obtained from the integration of the loss term in the Boltzmann equation. Indeed, this integral computes the rate at which a particle of \textit{any} velocity changes velocity due to collision, which is by definition the collision frequency \cite{visco2008collisional}.

Eqs.~\eqref{eq: MeanValues} and \eqref{eq: energy change microscopic} lead to the following equation for the steady state temperature used in the main text:
\begin{equation}
    0=\frac{\omega}{2}(m\Delta^2+\alpha\Delta\sqrt{\pi m T}-T(1-\alpha^2))-2\gamma (T-T_b).
    \label{eq: appendix temperature steady state}
\end{equation}

\section{Details concerning the \texorpdfstring{$\Delta + \gamma$}{Δ+γ} model }
\label{sec: appendix details delta}

In the $\Delta+\gamma$ model, $T_b = 0$, therefore $\mathcal{G}$ reduces to:
\begin{equation}
    \phi g^+\equiv\mathcal{G}^{\Delta+\gamma}=\dfrac{\sigma \gamma \sqrt{\pi m} T}{2\sqrt{T}(m\Delta^ 2 + \alpha\Delta \sqrt{\pi m T}-(1-\alpha^2) T)},
    \label{eq: def of G delta}
\end{equation}
and $\tilde p^{\Delta+\gamma}$ reads:
\begin{equation}
    \tilde p^{\Delta+\gamma}= T +  \mathcal{G}^{\Delta+\gamma}(T)\left((1+\alpha)T+ 2\Delta\sqrt{\pi m T}\right).
    \label{eq: p tilde delta}
\end{equation}

In order to know whether $T_s$ is larger or smaller than $T_l$ using the inequality $\tilde p^{\Delta + \gamma}(T_s)<\tilde p^{\Delta + \gamma}(T_l)$, we must know the variations of $\tilde p$. Direct computation of the derivative of $\tilde p^{\Delta + \gamma}$ with respect to $T$ leads to:
\begin{equation}
    \begin{split}
        \dfrac{\dd \tilde p^{\Delta+\gamma}}{\dd T} =&~ 1 + \left( 1 + \alpha + \Delta \sqrt{\pi m/T} \right) \mathcal{G}^{\Delta+\gamma} \\
        &+\left( (1+\alpha)T + 2 \Delta \sqrt{\pi m T} \right) \frac{\dd \mathcal{G}^{\Delta+\gamma}}{\dd T},
    \end{split}
\end{equation}
using $m\Delta^ 2 + \alpha\Delta \sqrt{\pi m T}-(1-\alpha^2) T=4\gamma T/\omega>0$ (Eq.~\eqref{eq: appendix temperature steady state}), we can show that $\frac{\dd \mathcal{G}^{\Delta + \gamma}}{\dd T}>0$ and $\mathcal{G}^{\Delta + \gamma}>0$ from which we obtain that $\dd \tilde p^{\Delta+\gamma}/\dd T>0$. This proves that $\tilde p^{\Delta+\gamma}$ is a continuous and increasing function of the relevant temperatures. 

Using the monotonous increase of $\tilde p^{\Delta+\gamma}$, we finally arrive to our result:
\begin{equation}
    \tilde p^{\Delta+\gamma}(T^{\Delta+\gamma}_s) < \tilde p^{\Delta+\gamma}(T^{\Delta+\gamma}_l)\Rightarrow T^{\Delta+\gamma}_s < T^{\Delta+\gamma}_l,    
\end{equation}
which shows that the solid is colder than the liquid at coexistence in the $\Delta+\gamma$ model.

\section{Details concerning the GLM \label{sec: appendix details langevin}}

In the GLM, $\Delta = 0$, therefore $\mathcal{G}$ simplifies to:
\begin{equation}
    \phi g^+\equiv \mathcal{G}^\text{GLM}=\dfrac{\sigma \gamma \sqrt{\pi m} (T_b-T)}{2(1-\alpha^2) T^{3/2}},
    \label{eq: appendix def of G langevin}
\end{equation}
and $\tilde p^\text{GLM}$ reads:
\begin{equation}
    \begin{split}
    \tilde p^\text{GLM}&= T\left(1 +  \mathcal{G}^\text{GLM}(T)(1+\alpha)\right)\\
    &=T_b\tilde T\left(1 + \Lambda (1-\tilde T)\tilde T^{-3/2}\right),
    \label{eq: appendix p tilde langevin}
    \end{split}
\end{equation}
with $\tilde T = T/T_b$ and $\Lambda = \frac{\gamma\sigma\sqrt{\pi m/T_b}}{2(1-\alpha)}>0$ a dimensionless parameter such that, when molecular chaos holds and velocities follow a gaussian distribution, $\Lambda\to \infty$ corresponds to an equilibrium limit. 

\begin{figure}[t]
\includegraphics[width=0.98\columnwidth,clip=true]{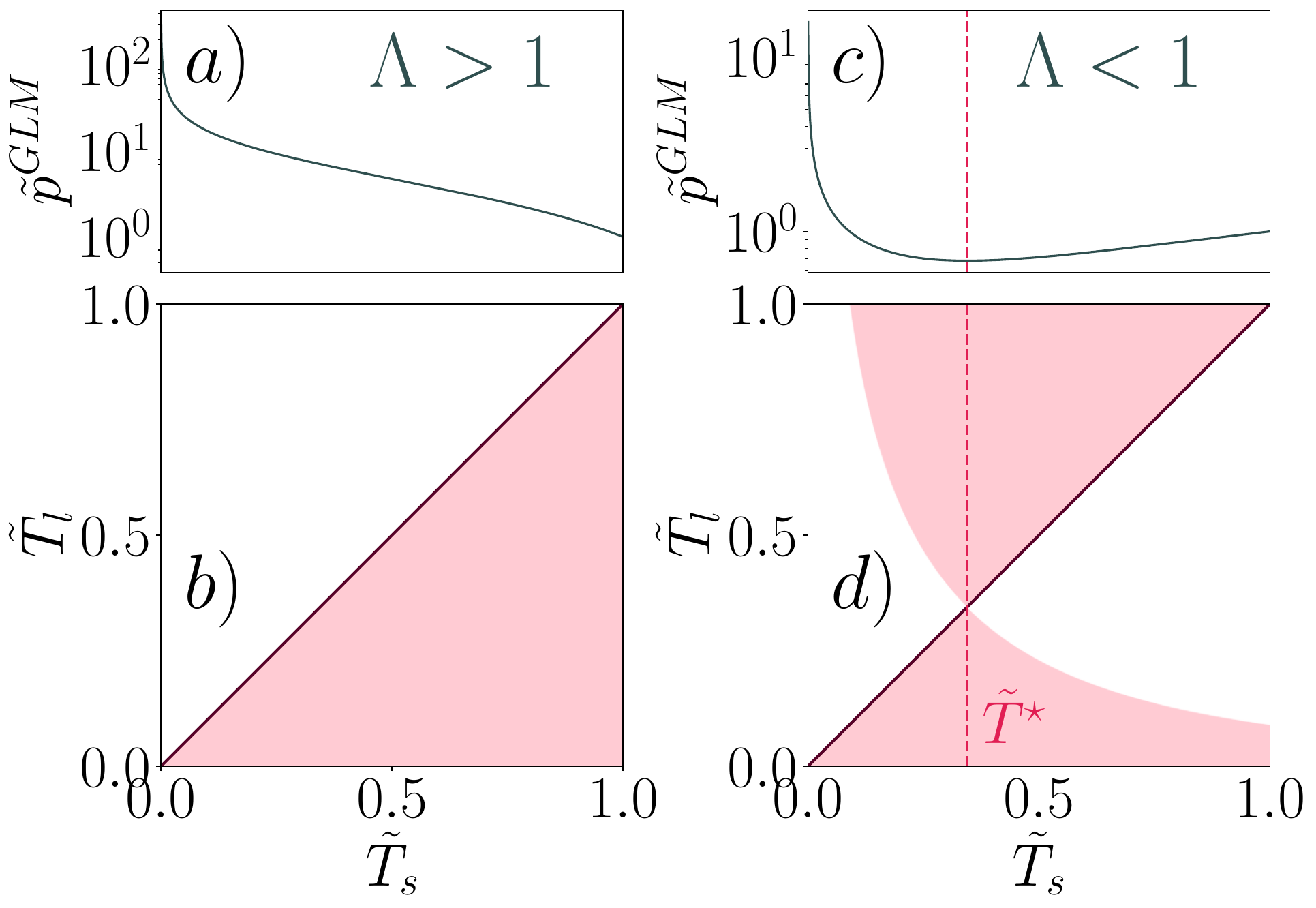}
\centering
\caption{ Typical behavior of $\tilde p^\text{GLM}$ and  $\tilde p^\text{GLM}(T_s)<\tilde p^\text{GLM}(T_l)$ for $\Lambda > 1$ and $\Lambda <1$. a) and c) Evolution of $\tilde p^\text{GLM}$ as a function of $\tilde T$. For $\Lambda > 1$, $\tilde p$ is purely decreasing while for $\Lambda > 1$, $\tilde p^\text{GLM}$ has a change of variation at $\tilde T^\star$. b) and d) The corresponding regions which verify $\tilde p^\text{GLM}(T_s)<\tilde p^\text{GLM}(T_l)$. The shaded regions are the region in which the inequality holds. For $\Lambda > 1$, we obtain $\tilde T_s>\tilde T_l$. However, for $\Lambda < 1$, both scenario are possible. Hence, determining whether $\tilde T_s>\tilde T_l$ requires  estimates for $\phi g^+$.}  \label{fig: appendix granular langevin}
\end{figure}

Since the bath is the only source of energy injection in the system, physical temperatures must respect $T < T_b$ and hence $\tilde T < 1$. We can therefore restrict our attention to the behavior of $\tilde p^\text{GLM}$ at $\tilde T<1$. In order to know whether $T_s$ is larger or smaller than $T_l$ using the inequality $\tilde p^\text{GLM}(T_s^\text{GLM})<\tilde p^\text{GLM}(T_l^\text{GLM})$, we must know the variations of $\tilde p^\text{GLM}$. Direct computation of the derivative of $\tilde p^\text{GLM}$ with respect to $T$ leads to:
\begin{equation}
    \dfrac{\dd \tilde p^\text{GLM}}{\dd T}= T_b\left(1- \dfrac{\Lambda}{2}(1+\tilde T)\tilde T^{-3/2}\right).
    \label{eq: derivative langevin}
\end{equation}

Eq.~\eqref{eq: derivative langevin} implies that $\tilde p^\text{GLM}$ is decreasing from 0 to $\tilde T^\star$ and then increasing from $\tilde T^\star$ to $\infty$, where $\tilde T^\star$ is the unique real root of Eq.~\eqref{eq: derivative langevin} given by a third order polynomial in $\tilde T^{1/2}$:
\begin{equation}
     2(\tilde T^\star)^{3/2}- \Lambda(1+\tilde T^\star)=0.
     \label{eq: andrea polynomial}
\end{equation}
We now perform an asymptotic expansion around $\Lambda = 1$ where the solution $\tilde T^\star=1$ of the polynomial Eq.~\eqref{eq: andrea polynomial} is known:
\begin{equation}
    \begin{split}
        \Lambda &= 1 + \varepsilon\\
        \tilde T^{\star} &= 1 + \sum_{i=1}^{\infty} a_i\varepsilon^i.
    \end{split}
    \label{eq: asymptotic expansion}
\end{equation}
Inserting Eq.~\eqref{eq: asymptotic expansion} in Eq.~\eqref{eq: andrea polynomial} and solving order by order leads the following approximation for the root of Eq.~\eqref{eq: derivative langevin}:
\begin{equation}
    \tilde T^{*} = 1 + (\Lambda-1) + \dfrac{1}{8}(\Lambda - 1)^2 + \mathcal{O}\left((\Lambda-2)^3\right).
\end{equation}
For $\Lambda > 1$, as it is clear that the exact root $\tilde T^{\star}(\Lambda)$ is monotonically increasing, it must satisfies $\tilde T^\star>1$, therefore, for the physically accessible temperatures ($\tilde T<1$), $\tilde p^\text{GLM}$ is decreasing which implies:
\begin{equation}
    \begin{split}
    \tilde{p}^\text{GLM}(T^\text{GLM}_s) &< \tilde{p}^\text{GLM}(T^\text{GLM}_l) \\
    &\Downarrow ~ \Lambda > 1 \\
    T^\text{GLM}_s &> T^\text{GLM}_l.
    \end{split}
\end{equation}
This behavior of $\tilde p^\text{GLM}$ is exemplified in Fig.~\ref{fig: appendix granular langevin}a) and the corresponding constraints on $\tilde T_s$ and $\tilde T_l$ are given by Fig.~\ref{fig: appendix granular langevin}b) where the shaded area are the region in which the inequality Eq.~\eqref{eq: criterion pressure} is verified.

In contrast, when $\Lambda < 1$, $\tilde p^\text{GLM}$ changes from being decreasing to increasing at a value of $\tilde T^{*}$ between 0 and 1. Hence, the constraint on the pressure alone does not determine which phase is hotter. This is exemplified in Fig.~\ref{fig: appendix granular langevin}c) and d). However in this limit, the system was typically always observed to crystallize via a continuous transition, rather than through a phase coexistence, in numerical simulations (see Appendix.~\ref{sec: appendix continuous}).

\bibliographystyle{unsrt}

\end{document}